\input amstex
\documentstyle{amsppt}
\def\const{\operatorname{const}}
\def\grad{\operatorname{grad}}
\topmatter
\title
On the solution of normality equations \\
for the dimension $n\geq 3$.
\endtitle
\author
Andrey Yu. Boldin, Ruslan A. Sharipov
\endauthor
\thanks
The work is done under financial support of European Fund
INTAS (project \#93-47, head of project: Pin\-chuk~S.~I.)
and Russian Fund of Fundamental Investigations (project \#96-01-00273,
head of project: Sultanaev~Y.~T.).
\endthanks
\address
Bashkir State University,
Frunze str., 32, Ufa,
Russia, 450074
\endaddress
\email
root\@bgua.bashkiria.su
\endemail
\abstract
The normality equations for the Newtonian dynamical systems
on an arbitrary Riemannian manifold of the dimension $n \geq 3$
are considered. Locally the solution of such equations reduces
to three possible cases: in two of them the solution is written
out explicitly, and in the third case the equations of normality
are reduced to an ordinary differential equation of the second
order. Some new examples of explicit solutions of normality
equations are constructed.
\endabstract
\endtopmatter
\document
\head
1. Introduction.
\endhead
The concept of dynamical systems that admit the normal shift was
introduced in \cite{1,2}. These are the systems which describe
the motion of a mass point according to the Newton's second law
$m\,\ddot\bold r=\bold F(\bold r,\dot\bold r)$ and which satisfy
some specific geometrical condition that permits them to
implement the normal shift of any hypersurface along their
trajectories. Such geometrical condition was called {\it the
condition of normality}. In \cite{1,2} this geometrical condition
of normality was brought to the form of a system of partial
differential equations for the force field $\bold F(\bold r,\dot
\bold r)$ of the dynamical system. These equations were called
{\it the equations of normality}.\par
      In \cite{3,4} the equations of normality were generalized
for the case of Newtonian dynamical systems on an arbitrary Riemannian
manifold. Note that in this case the normality equations were
written in covariant form. Therefore one can apply the methods
of differential geometry to these equations.\par
     The geodesic flows of the metrics that is conformally equivalent
to an original metric of the Riemannian manifold form a wide subclass
of the dynamical systems accepting the normal shift. The problem
of coincidence of trajectories of a dynamical system with the
trajectories of such geodesic flow was called {\it the problem
of metrizability}. It was considered in \cite{5,6} where an explicit form of
the force field for all metrizable dynamical system that admit the normal
shift was obtained.\par
     The symmetry analysis of the equation of normality was started
in \cite{7} where all classical symmetries of these equations in
two-dimensional spatially homogeneous case were found and the
corresponding self-similar (invariant) solutions were con\-struct\-ed.
As a result of the symmetry analysis some examples of non-trivial
(non-metrizable) dynamical systems accepting the normal shift in
two-dimensional case were found. The importance of the search for
such dynamical systems especially for the dimension $n \geq 3$ was
pointed out to authors by academician A.~T.~Fomenko. Some simplest
examples of non-metrizable systems were constructed in \cite{8},
however the systematical search for such systems has yet to be
undertaken.\par
     The main goal of the present article is the complete local
analysis of normality equations for the dimension $n \geq 3$ and
the construction of new non-metrizable dynamical systems that admit
the normal shift.\par
\head
2. The equations of normality.
\endhead
     Let $M$ be a $n$-dimensional Riemannian manifold with the metric
tensor $g_{ij}$. Let's denote by $x^1,\ldots,x^n$ the local coordinates
of some point on $M$ and by $v^1,\ldots,v^n$ the coordinates
of a tangent vector at this point. The equations of Newtonian dynamical
system describing the motion of the point of unit mass $m=1$ have
the following form:
$$
\xalignat 2
&\dot x^i=v^i,
&
&\nabla_t v^i=F^i(\bold x,\bold v).
\tag2.1
\endxalignat
$$
The normality equations for the force field $\bold F$ of the
dynamical system \thetag{2.1} are written as two systems of
differential equations:
$$
\align
&\left\{
\aligned
&\bigl(v^{-1} F_i+\tilde\nabla_i(F^k N_k)\bigr)P_q^i = 0\\
\vspace{1ex}
&(\nabla_i F_k+\nabla_k F_i-2 v^{-2}F_i F_k) N^k P_q^i+\\
&+v^{-1}(\tilde\nabla_k F_i F^k
 -\tilde\nabla_k F^r N^k N_r F_i) P_q^i = 0
\endaligned
\right.\tag2.2 \\
\vspace{1ex}
&\left\{
\aligned
&(P^k_i P^q_j-P^q_i P^k_j)
\left( N^r \frac{\tilde\nabla_r F_k}{v} F_q-
\nabla_q F_k\right)=0\hskip-2ex\\
&P^k_i \tilde\nabla_k F^q P^j_q =
\frac{P^k_r \tilde\nabla_k F^q P^r_q}{n-1} P^j_i
\endaligned
\right.\tag2.3
\endalign
$$
The systems of equations \thetag{2.2} and \thetag{2.3} were derived
separately in \cite{3} and \cite{4}. The first system was called
{\it the equations of weak normality}, and the second one was called
{\it the additional normality equations}. The equations of
normality \thetag{2.2} and \thetag{2.3} are differential equations in
the covariant derivatives $\nabla$ and $\tilde\nabla$ in the expanded
algebra of tensor fields. A tensor field from the expanded algebra is
distinguished from an ordinary tensor field on $M$ by the doubled set
of arguments: it depends on a point $\bold x \in M$ as well as a
tangent vector $\bold v$ at this point. Therefore there are two types of
covariant derivatives in the expanded algebra: {\it the spatial gradient}
$\nabla$ and {\it the velocity gradient} $\tilde\nabla$. The calculation
of components of the covariant derivative $\tilde\nabla$ for a tensor
field $\bold U$ of $(r,s)$--type is simply the differentiation
with respect to the components of velocity:
$$
\tilde\nabla_iU^{i_1\ldots i_r}_{j_1\ldots j_s}=
\frac{\partial U^{i_1\ldots i_r}_{j_1\ldots j_s}}
{\partial v^i}.
\tag2.4
$$
There are no connection components $\Gamma^k_{ij}$ in \thetag{2.4}.
Therefore the covariant derivative $\tilde\nabla$ can be defined
on a manifold without any metric. The components of the covariant
derivative $\nabla$ are calculated in more complicated way:
$$
\aligned
&\nabla_iU^{i_1\ldots i_r}_{j_1\ldots j_s}=
\frac{\partial U^{i_1\ldots i_r}_{j_1\ldots j_s}}{\partial x^i}+
\sum^r_{m=1}\sum^n_{q_m=1}U^{i_1\ldots q_m\ldots i_r}_{j_1\ldots
j_s}\,\Gamma^{i_m}_{i\,q_m}-\\
&-\sum^s_{m=1}\sum^n_{q_m=1}U^{i_1\ldots i_r}_{j_1\ldots q_m\ldots
j_s}\,\Gamma^{q_m}_{i\,j_m}-\sum^n_{p=1}\sum^n_{q=1}
\frac{\partial U^{i_1\ldots i_r}_{j_1\ldots j_s}}{\partial v^p}
\,\Gamma^p_{i\,q}\,v^q.
\endaligned
\tag2.5
$$\par
      The equations \thetag{2.2} and \thetag{2.3} are the equations
with respect to the force vector $\bold F$. But apart from the components
of the force vector, in \thetag{2.2} and \thetag{2.3} there are also the
components of tensor fields $\bold N$ and $\bold P$, and a scalar field $v$.
These are given parameters in the normality equations: $v$ is the field of
velocity modulus $v=|\bold v|$; $\bold  N$  is  the  unit  vector
directed
along $\bold v$ and $\bold P$ is the orthogonal projector on the hyperplane
that is perpendicular to the vector $\bold v$:
$$
\xalignat 2
&N^i=v^i/v,
&
&P^i_j=\delta^i_j-N^i\,N_j.
\tag2.6
\endxalignat
$$\par
      Both covariant derivatives $\nabla$ and $\tilde\nabla$ of
a metric tensor are equal to zero. This makes calculations in the expanded
algebra similar to calculations in the ordinary algebra of tensor
fields on $M$. The covariant derivatives of given fields
$v$, $\bold N$ and $\bold P$ are calculated from the formulae
\thetag{2.4} and \thetag{2.5}.\par
\head
3. A scalar substitution.
\endhead
Consider the first equation from \thetag{2.2}. Let us denote $A=F^k N_k$.
The scalar field $A$ from the expanded algebra has the sense of the
orthogonal projection of the force $\bold F$ on the direction of
the velocity vector. So, the force field $F$ can be written in the
form: $\bold F=A\,\bold N+\tilde\bold F$. Substituting this expansion
into the first equation from \thetag{2.2}, we find:
$$
v^{-1}\,\tilde F_i\,P^i_q+\tilde\nabla_i A\,P^i_q=0.
\tag3.1
$$
Here we've used the obvious equality $P^i_q\,N_i=0$, which follows
from \thetag{2.6}, and the fact that $\bold N$ is a unit vector. From
the orthogonality of the vectors $\tilde\bold F$ and $\bold N$ we have
$\tilde F_i\,P^i_q=\tilde F_q$. So, taking into account everything told
above, from \thetag{3.1} we obtain:
$$
F_q=A\,N_q-v\,P^i_q\,\tilde\nabla_i A.
\tag3.2
$$
From \thetag{3.2} we see that all components of the force field $\bold F$
are defined by the scalar field $A$.\par
      Let's substitute \thetag{3.2} into the second equation of weak
normality \thetag{2.2}. It leads to quite cumbersome, but not complicated
computations. As a result of these computations we obtain:
$$
\aligned
P^i_q\,(\nabla_iA&+v\,P^{ks}\,\tilde\nabla_sA\,
\tilde\nabla_k\tilde\nabla_iA-\\
&-v\,N^k\,\nabla_k\tilde\nabla_iA-
N^k\,A\,\tilde\nabla_k\tilde\nabla_iA)=0.
\endaligned
\tag3.3
$$
In order to derive \thetag{3.3} we've used the following relationships:
$$
\xalignat 2
\nabla_iv&=0,
&\tilde\nabla_iv&=N_i,
\tag3.4\\
\nabla_iN^k&=0,
&\tilde\nabla_iN^k&=v^{-1}\,P^k_i,
\tag3.5\\
\nabla_i P^k_q&=0,
&\tilde\nabla_iP^k_q&=-v^{-1}\,(P_{iq}\,N^k+P^k_i\,N_q),
\tag3.6
\endxalignat
$$
that are obtained by direct computations taking into account the
relationships \thetag{2.4}, \thetag{2.5} and \thetag{2.6}.\par
      Now let us substitute \thetag{3.2} into the additional equations of
normality \thetag{2.3} and use once more the relationships \thetag{3.4},
\thetag{3.5} and \thetag{3.6} in combination with $P^i_q\,N_i=0$ and
$P^i_q\,P^q_k=P^i_k$. From the first equation \thetag{2.3} we derive
$$
\aligned
P^k_i\,P^q_j\,(&N^r\,\tilde\nabla_r\tilde\nabla_kA\,
\tilde\nabla_qA+\nabla_q\tilde\nabla_kA-\\
&-N^r\,\tilde\nabla_r\tilde\nabla_qA\,\tilde\nabla_kA-
\nabla_k\tilde\nabla_qA)=0.
\endaligned
\tag3.7
$$
After the substitution \thetag{3.2} into the second equation \thetag{2.3}
we obtain:
$$
P^k_i\,\tilde\nabla_k\tilde\nabla_qA\,P^{qj}=
\frac{P^{kq}\,\tilde\nabla_k\tilde\nabla_qA}{n-1}\,P^j_i,
\tag3.8
$$
where the number $n$ is a dimension of the manifold $M$. It's convenient
to rewrite the last equation in the following form:
$$
P^k_i\,\tilde\nabla_k\tilde\nabla_qA\,P^{qj}=\lambda\,P^j_i,
\tag3.9
$$
where $\lambda=\lambda(\bold x,\bold v)$ is a function on the tangent
bundle $TM$. It is the very form this equation was initially derived in
\cite{4}.\par
\head
4. Fiber spherical coordinates.
\endhead
The equations \thetag{3.3}, \thetag{3.7} and \thetag{3.9}
form the complete list of the equations of normality after
the scalar substitution \thetag{3.2}. The last equation from
this list has an important difference from the other ones. If
we write out the covariant derivatives $\tilde\nabla_k\tilde\nabla_qA$,
in explicit form, from \thetag{3.9} we obtain:
$$
P^k_i\,P^q_j\,\frac{\partial^2A}{\partial v^k\,\partial v^q}
=\lambda\,P_{ij}.
\tag4.1
$$
The equations \thetag{4.1} contain only the derivatives of $A$ with respect
to the coordinates $v^1,\ldots,v^n$ in the fiber of the tangent bundle $TM$.
The fiber is a $n$-dimensional linear vector space with the metric $g_{ij}$
which is constant within the fiber.\par
     Let's fix some point $P$ on the manifold $M$. The conditions
$v=|\bold v|=\const$ decompose the fiber of $TM$ over $P$ into the
union of non-intersecting spheres $S^{n-1}$ with the radii $r=v$.
Let $u^1,\ldots,u^{n-1}$ be local coordinates on the unit sphere
$S^{n-1}$ and let $\bold v=\bold N(u^1,\ldots,u^{n-1})$ be the parametrical
equation of the unit sphere in the fiber of $TM$ over the point $P$.
Then the relationships
$$
\matrix
v^1=u^n\cdot N^1(u^1,\ldots,u^{n-1}),\\
\hdotsfor 1\\
v^n=u^n\cdot N^n(u^1,\ldots,u^{n-1})
\endmatrix
\tag4.2
$$
define the change of Cartesian coordinates $v^1,\ldots,v^n$ for
the spherical coordinates $u^1,\ldots,u^n$, where $u^n=|\bold v|$.
The choice of coordinates \thetag{4.2} in each fiber of $TM$ defines
{\it the fiber spherical coordinates} on the tangent bundle $TM$. The
dependence of the choice of such coordinates upon the point $P$ can be
made smooth, however for analysis of the equations \thetag{4.1} this does
not matter. Therefore we'll consider the relationships \thetag{4.2}
taking the point $P$ to be fixed.\par
      Let's transform the equations \thetag{4.1} by making the change of
variables \thetag{4.2}. The metric $g_{ij}$ is constant within
the fiber and the derivatives $\tilde\nabla_i=\partial/\partial v^i$
are covariant derivatives in the metric $g_{ij}$. The equations
\thetag{4.1} are tensorial equations, therefore it's sufficient to
recalculate the components of tensors $P^k_i$, $P^q_j$, $g_{ij}$
and the derivatives $\tilde\nabla_i$ in the new coordinates.\par
      Note that according to \thetag{4.2} we have $\partial v^i/\partial
u^n=N^i$. Therefore vector $\bold N$ is the $n$-th coordinate vector in the
spherical coordinates: $\bold N=(0,\ldots,0,1)$. So, for the components
of the projector $\bold P$ in spherical coordinates we have
$$
\xalignat 2
&\quad P^k_i=
\Vmatrix
1      & \hdots & 0      & 0      \\
\vdots & \ddots & \vdots & \vdots \\
0      & \hdots & 1      & 0      \\
\vspace{1ex}
0      & \hdots & 0      & 0
\endVmatrix,
&&P_{ij}=
\Vmatrix
g_{11}     & \hdots & g_{1\,n-1}   & 0      \\
\vdots     & \ddots & \vdots       & \vdots \\
g_{n-1\,1} & \hdots & g_{n-1\,n-1} & 0      \\
\vspace{1ex}
0      & \hdots & 0      & 0
\endVmatrix.
\tag4.3
\endxalignat
$$
In view of what we said above, in the spherical coordinates the
equation \thetag{4.1} has the initial form \thetag{3.9}. But
the components of the connection $\vartheta^k_{ij}$ in
spherical coordinates $u^1,\ldots,u^n$ are nonzero. For $i,j,k$
in the range $1\leq i,j,k\leq n-1$ the quantities
$\vartheta^k_{ij}$ do coincide with the components of the metric connection
on the spheres $|\bold v|=\const$. The other components are calculated
explicitly:
$$
\xalignat 3
&\vartheta^n_{nn}=0,
&&\vartheta^n_{in}=\vartheta^n_{ni}=0,
&&\vartheta^i_{nn}=0,\\
\vspace{-1ex}
&&&&&\tag4.4\\
\vspace{-2ex}
&\vartheta^n_{ij}=-\frac{g_{ij}}{v},
&&\vartheta^i_{jn}=\vartheta^i_{nj}=\frac{\delta^i_j}{v}.
&&\\
\endxalignat
$$
By virtue of the structure of the matrices \thetag{4.3} for $i=n$
or $j=n$ the equations \thetag{3.9} are fulfilled identically. Taking
into account \thetag{4.4} we can write the rest non-trivial equations
as the equations in the covariant derivatives $\bar\nabla$ on the spheres
$|\bold v|=\const$:
$$
\bar\nabla_i\bar\nabla_jA=\left(\lambda+\frac{1}{v}\,
\frac{\partial A}{\partial u^n}\right)\,g_{ij}.
\tag4.5
$$
Since the function $\lambda$ is an indeterminate parameter in
\thetag{4.5}, it's convenient to introduce another parameter $\mu$ and
to write \thetag{4.5} more simply:
$$
\bar\nabla_i\bar\nabla_jA=\mu\,g_{ij}.
\tag4.6
$$\par
      Consider a differential consequence from the equation \thetag{4.6}.
Let us write \thetag{4.6} in the form $\bar\nabla_j\bar\nabla_kA=
\mu\,g_{jk}$ then apply the operator $\bar\nabla_i$ to both sides of
this equation and alternate the obtained equation with respect to
the indices $i$ and $j$:
$$
\align
&(\bar\nabla_i\bar\nabla_j-\bar\nabla_j\bar\nabla_i)
\bar\nabla_kA=\bar\nabla_i\mu\,g_{jk}-\bar\nabla_j\mu\,
g_{ik}.\\
&-\bar R^s_{kij}\,\bar\nabla_sA=\bar\nabla_i\mu\,g_{jk}-
\bar\nabla_j\mu\,g_{ik}.
\tag4.7
\endalign
$$
Here $R^s_{kij}=v^{-2}\,(\delta^s_i\,g_{jk}-\delta^s_j\,
g_{ik})$ is the curvature tensor of the sphere $S^{n-1}$ given
by the equation $v=|\bold v|=\const$.\par
      Let's contract the above equation \thetag{4.7} with the
metric $g^{jk}$. It leads to the following relationship:
$$
\frac{n-2}{v^2}\,\bar\nabla_iA=(n-2)\,\bar\nabla_i\mu.
$$
Note that the additional equations of normality \thetag{2.3} arise
in the dimension $n\geq 3$ only, therefore in the obtained equations
$n-2\neq 0$. Hence
$$
-\nabla_iA=v^2\,\nabla_i\mu.
\tag4.8
$$
Let's substitute the last expression for $\nabla_iA$ into the
equations \thetag{4.6}. It leads to the following differential
equations for the functional parameter $\mu$:
$$
\bar\nabla_i\bar\nabla_j\mu=-v^{-2}\,g_{ij}\,\mu.
\tag4.9
$$
\proclaim{Theorem 4.1} The general solution of the system of
equations \thetag{4.9} on the sphere $|\bold v|=\const$ has the form:
$$
\mu=\sum^n_{i=1} m_i\,N^i(u^1,\ldots,u^n),
\tag4.10
$$
where $m_1,\ldots,m_n$ are some arbitrary constants and $N^1,\ldots,N^n$
are the components of the vector $\bold N$.
\endproclaim
\demo{Proof} Because the equations \thetag{4.9} and the expression
\thetag{4.10} are covariant tensorial equalities it's convenient to
check the statement of the theorem for some particular choice of coordinates
$u^1,\ldots,u^n$ on the sphere $|\bold v|=\const$. Let's render concrete
the change of variables \thetag{4.2}, taking the metric $g_{ij}$ to be
brought to the unitary form $g_{ij}=\delta_{ij}$ in the coordinates
$v^1,\ldots,v^n$ in the fiber of $TM$ over the considered fixed point $P$:
$$
\matrix
\format\l&\,\l&\l\\
v^1&=v\,\sin(u_{n-1})\,\sin(u_{n-2})\cdot\ldots\cdot\sin(u_3)\,
&\sin(u_2)\,\sin(u_1),\\
v^2&=v\,\sin(u_{n-1})\,\sin(u_{n-2})\cdot\ldots\cdot\sin(u_3)\,
&\sin(u_2)\,\cos(u_1),\\
v^3&=v\,\sin(u_{n-1})\,\sin(u_{n-2})\cdot\ldots\cdot\sin(u_3)\,
&\cos(u_2),\\
\hdotsfor 2\\
\vspace{1ex}
v^{n-1}&=v\,\sin(u_{n-1})\,\cos(u_{n-2}),\\
v^n&=v\,\cos(u_{n-1}).\\
\endmatrix
\tag4.11
$$
In spherical coordinates $u^1,\ldots,u^n$ defined by the relationships
\thetag{4.11} the metric $g_{ij}$ is diagonal, moreover $g_{ii}>0$
and the quantities $H_i=\sqrt{g_{ii}}$ are called {\it the Lame
coefficients}:
$$
\matrix
\format\l&\,\l&\l\\
H_1&=v\,\sin(u_{n-1})\,\sin(u_{n-2})\cdot\ldots&\cdot\sin(u_3)\sin(u_2),\\
H_2&=v\,\sin(u_{n-1})\,\sin(u_{n-2})\cdot\ldots&\cdot\sin(u_3),\\
\hdotsfor 2\\
\vspace{1ex}
H_{n-2}&=v\,\sin(u_{n-1}),\\
H_{n-1}&=v,\\
H_n&=1.\\
\endmatrix
\tag4.12
$$
The components of the metrical connection $\vartheta^k_{ij}$ are
defined by the Lame coefficients \thetag{4.12} in the following way:
$$
\vartheta^k_{ij}=\frac{1}{H_k}\,\frac{\partial H_k}{\partial u^j}
\,\delta_{ik}+\frac{1}{H_k}\,\frac{\partial H_k}{\partial u^i}\,
\delta_{jk}-\frac{H_i}{(H_k)^2}\,\frac{\partial H_i}{\partial u^k}\,
\delta_{ij}.
$$
Most of the components of the connection $\vartheta^k_{ij}$ are zero.
The calculation of the other ones makes possible to write \thetag{4.9}
explicitly. As a result we obtain the equations of two types.
The equations of the first type correspond to the case $i<j\leq n-1$ in
\thetag{4.9}:
$$
\mu_{ij}=\frac{\cos(u_j)}{\sin(u_j)}\,\mu_i.
\tag4.13
$$
Here $\mu_i$ and $\mu_{ij}$ are the partial derivatives of $\mu$ of
the first and the second order with respect to the corresponding
variables $u^i$ and $u^j$. The equations of the second type correspond to
the case $i=j\leq n-1$:
$$
\mu_{ii}+\sum^{n-1}_{k=i+1}\frac{\cos(u^k)}{\sin(u^k)}\,
\left(\,\shave{\prod^{k}_{p=i+1}}\sin^2(u^p)\right)\,\mu_k=
-\left(\,\shave{\prod^{n-1}_{p=i+1}}\sin^2(u^p)\right)\,\mu.
\tag4.14
$$
Note that, when $i=n-1$, the equation \thetag{4.14} is reduced to
the form $\mu_{ii}=-\mu$.\par
      Let us make the further proof of the theorem by induction on $n$.
We take, as the base of induction, the case $n=2$. In the initial
geometrical problem the additional equations of normality arise only
for the dimension $n>2$. Therefore the equations \thetag{4.6} are also
absent. The relationship \thetag{4.8}, which lets us bring \thetag{4.6}
to \thetag{4.9}, was also derived for $n>2$. But once the equations
\thetag{4.9} are already derived, their extrapolation for the case $n=2$
has no contradiction.\par
      So, when $n=2$, the systems \thetag{4.13} and \thetag{4.14} are reduced
to the equation: $\mu_{11}=-\mu$. The general solution of this equation is:
$$
\mu(u^1)=m_1\,\sin(u^1)+m_2\,\cos(u^1),
\tag4.15
$$
where $m_1$ and $m_2$ are constants. In the case $n=2$ the relationships
\thetag{4.11} have the form:
$$
\xalignat 2
&v^1=v\,N^1=v\,\sin(u_1),
&
&v^2=v\,N^2=v\,\cos(u_1).
\tag4.16
\endxalignat
$$
Comparison of \thetag{4.15} with \thetag{4.10} and taking into account
\thetag{4.16} show that the statement of theorem 4.1 is fulfilled for
the case $n=2$.\par
      Now let $n>2$. Consider the last equation from \thetag{4.14} with
$i=n-1$. It has the form $\mu_{ii}=-\mu$. Hence
$$
\mu=\tilde \mu\,\sin(u^{n-1})+m_n\,\cos(u^{n-1}),
\tag4.17
$$
where $\tilde \mu$ and $m_n$ are parameters that in general case
depend on $u^1,\ldots,u^{n-2}$. Let's substitute \thetag{4.17}
into the equations \thetag{4.13}, setting $j=n-1$. As a result
the contribution of the first summand from \thetag{4.17} is canceled
and we obtain the following relationships for $m_n$:
$$
\frac{\partial m_n}{\partial u^i}=0\text{ \ for all \ }
i=1,\ldots,n-2.
$$
Hence $m_n=\const$. Therefore the substitution of \thetag{4.17} into the
rest equations \thetag{4.13} and \thetag{4.14} leads to the equations for
the parameter $\tilde \mu$, which have the same form as \thetag{4.13} and
\thetag{4.14}, but with $n-1$ in place of $n$. It means that one can
apply the assumption of induction. As a result we obtain:
$$
\tilde\mu=m_1\,\left(\,\shave{\prod^{n-2}_{p=1}}\sin(u^p)\right)+
\sum^{n-1}_{k=2}m_k\,\left(\,\shave{\prod^{n-2}_{p=k}}\sin(u^p)\right)
\,\cos(u^{p-1}),
\tag4.18
$$
where $m_1,\ldots,m_{n-1}$ are some constants. Now it remains to substitute
\thetag{4.18} into \thetag{4.17} and to compare each of the summands in the
obtained formula with \thetag{4.11}. From this comparison we see that
\thetag{4.18} and \thetag{4.10} are the same. So, the theorem is proved.
\qed\enddemo
     Now let's return to the relationship \thetag{4.8} and write it in the
form $\nabla_i(A+v^2\,\mu)=0$, from where we obtain:
$$
A+v^2\,\mu=\const.
\tag4.19
$$
Note that \thetag{4.19} and parameters $m_i$ from \thetag{4.10} are
constants within the separate spheres $|\bold v|=\const$ only. Therefore,
denoting $b_i=|\bold v|,\,m_i$, we can formulate the following statement.
\proclaim{Theorem 4.2} The scalar field $A(\bold x,\bold v)$ from the
extended algebra of tensor fields satisfying the equations of
normality \thetag{3.9} has the form:
$$
A=a(\bold x,|\bold v|)+|\bold v|\,b_i(\bold x,|\bold v|)
\,N^i,
\tag4.20
$$
where the scalar field $a$ and the covector field $\bold b$ from the expanded
algebra depend on the velocity modulus $v=|\bold v|$ in the fibers of the
tangent bundle $TM$.
\endproclaim
\head
5. The refinement of the scalar substitution.
\endhead
The obtained formula \thetag{4.20} completely defines the dependence of
the function $A$ upon velocity $\bold v$. The dependence of $A$ upon
coordinates should be refined by the substitution of \thetag{4.20} into
the equations \thetag{3.3} and \thetag{3.7}. Introduce the following
notations $a'=\partial a/\partial v$ and $a''=\partial^2 a/\partial v^2$.
It is not difficult to check that $a'$ and $a''$ are scalar fields from the
expanded algebra of tensor fields and they depend on velocity modulus
only. Moreover
$$
\xalignat 2
&\tilde\nabla_ia=a'\,N_i,
&
&\tilde\nabla_ia'=a''\,N_i.
\tag5.1
\endxalignat
$$
The derivatives $b'_k=\partial b_k/\partial v$ and $b''_k=\partial
b'_k/\partial v$ define two covector fields $\bold b'$ and $\bold b''$
from the expanded algebra of tensor fields. These fields also depend on
velocity modulus only and there are following relationships analogous to
\thetag{5.1}:
$$
\xalignat 2
&\tilde\nabla_ib_k=b'_k\,N_i,
&
&\tilde\nabla_ib'_k=b''_k\,N_i.
\tag5.2
\endxalignat
$$
Substituting \thetag{4.20} into the equations of normality \thetag{3.3}
and taking into account \thetag{5.1} and \thetag{5.2}, we obtain:
$$
P^i_q\,\bigl(\nabla_ia+a'\,b_i-b'_i\,a+v\,N^r\,\nabla_ib_r
-v\,N^r\,\nabla_rb_i\bigr)=0
\tag5.3
$$\par
      The equation \thetag{5.3} contains the covariant derivatives
$\nabla_ia$ and $\nabla_ib_r$. The formula \thetag{2.5} is
written in the variables $x^1,\ldots,x^n,v^1,\ldots,v^n$. But
for the fields $a$ and $\bold b$ the natural variables are
$x^1\ldots,x^n,v$, where $v=|\bold v|=\sqrt{g_{kq}\,v^k\,v^q}$.
While in the variables $x^1,\ldots,x^n,v^1,\ldots,v^n$ the velocity modulus
$v$ depends on $x^1,\ldots,x^n$ because $g_{kq}$ depends on $x^1,\ldots,x^n$.
Therefore
$$
\nabla_ia=\frac{\partial a}{\partial x^i}+
\frac{a'}{2v}\,\frac{\partial g_{kq}}{\partial x^i}\,v^k\,v^q-
\Gamma^k_{iq}\,\frac{a'}{v}\,v_k\,v^q.
$$
In view of concordance of the metric and connection two last
summands  in  this  expression   are   canceled.   An   analogous
cancellation happens
in the computation of $\nabla_ib_r$. For this reason the covariant
derivatives of the fields $a$ and $\bold b$ are calculated as if they
contain no dependence on $\bold v$:
$$
\xalignat 2
&\nabla_ia=\partial a/\partial x^i,
&&\nabla_ib_r=\partial b_r/\partial x^i-\Gamma^k_{ir}\,b_k.
\tag5.4
\endxalignat
$$
By virtue of \thetag{5.4} the covariant derivatives $\nabla_ia$ and
$\nabla_ib_r$ depend on the velocity modulus only.\par
      Let's replace the projector $P^i_q$ in \thetag{5.4} by the explicit
expression $P^i_q=\delta^i_q-N^i\,N_q$ from the formula \thetag{2.6}
$$
\aligned
\nabla_qa+a'\,b_q&-b'_q\,a+v\,(\nabla_qb_r-\nabla_rb_q)
\,N^r\\
&-N_q\,(\nabla_ra+a'\,b_r-b'_r\,a)\,N^r=0.
\endaligned
\tag5.5
$$
There are three groups of summands in the equation \thetag{5.5}. The first
three summands depend on the velocity modulus only. The other ones depend
on the modulus of the velocity vector and on its direction as well. The
direction of $\bold v$ is given by $\bold N$. One of these summands is linear
in $\bold N$, while another is quadratic with respect to $\bold N$.
The substitution of $\bold N$ by $-\bold N$ does not change the modulus of
the velocity vector. This substitution changes a sign of the summand that is
linear in $\bold N$ and the rest summands in \thetag{5.4}
remain unaltered. Therefore the summand linear with respect to $\bold N$
vanishes identically. Hence
$$
\nabla_qb_r-\nabla_rb_q=0.
\tag5.6
$$
The summand quadratic with respect to $\bold N$ contains the contraction
of the vector $\bold N$ and the covector with the components
$\nabla_ra+a'\,b_r-b'_r\,a$ which depend on $|\bold v|$ only.
While the modulus of the velocity vector is unchanged, the vector $\bold N$
can be made orthogonal to this covector. So, the summand quadratic in
$\bold N$ is also zero:
$$
\nabla_qa+a'\,b_q-b'_q\,a=0.
\tag5.7
$$
Conclusion: the equation of normality \thetag{3.3} for
the field $A$ of the form \thetag{4.20} is reduced to the equations
\thetag{5.6} and \thetag{5.7} for the fields $a$ and $\bold b$.\par
      The substitution of \thetag{4.20} into \thetag{3.7} add nothing new.
It leads to the equation that is the consequence of \thetag{5.6}. For this
reason \thetag{5.6} and \thetag{5.7} remain as unique basic
equations of normality for the fields $a$ and $\bold b$. Taking into
account the relationships \thetag{5.4} and symmetry of the connection
components $\Gamma^k_{ij}$ enables us to bring \thetag{5.6} to the form:
$$
\frac{\partial b_r}{\partial x^i}-\frac{\partial b_i}
{\partial x^r}=0.
\tag5.8
$$
The covector field $\bold b(\bold x,v)$ from the expanded algebra of
tensor fields depending only on the modulus of velocity can be treated
as the one-parameter family of $1$-forms on the manifold:
$\bold b=b_i\,dx^i$, where $v$ is a parameter. Then the equation \thetag{5.8}
is a closedness condition for each form from this family.\par
      \subhead Example 1\endsubhead Let $a\equiv 0$. Then the equation
\thetag{5.7} is fulfilled for any choice of $\bold b$. So, each one-
parameter family of closed $1$-forms corresponds to some Newtonian
dynamical system accepting the normal shift. The force field of such
system has the following form:
$$
F_q=|\bold v|\,b_i(\bold x,|\bold v|)\,(2\,N^i\,N_q-\delta^i_q).
\tag5.9
$$
Note that the simplest examples of the dynamical systems accepting
the normal shift from the papers \cite{1}, \cite{2} and \cite{8} are
systems of the form \thetag{5.9}.\par
      \subhead Example 2\endsubhead Let $\bold b\equiv 0$. Then the
equation \thetag{5.7} has the form: $\nabla_qa=0$. It means that
the function $a$ does not depend on a point of the manifold $M$, i.e.
$a=a(v)$. For the force field of the corresponding dynamical system
we have:
$$
F_q=a'(|\bold v|)\,N_q.
\tag5.10
$$
This is a trivial class of the dynamical systems that admit the normal
shift. The trajectories of the systems \thetag{5.10} coincide with
geodesics of the manifold $M$.\par
      Now consider the case $a\not\equiv 0$ and $\bold b\not\equiv 0$.
In this case the equation \thetag{5.7} is not trivial. Let's write it
in the following form:
$$
\left(\frac{\partial}{\partial x^q}+b_q\,\frac{\partial}
{\partial v}\right)\ln|a|=b'_q.
\tag5.11
$$
Let us set $\hat X_q=\partial/\partial x^q+b_q\,\partial/\partial v$
and compute the commutator of two such differential operators, taking
into account the relationships \thetag{5.8}:
$$
[\hat X_k,\,\hat X_q]=(b_k\,b'_q-b_q\,b'_k)\,
\partial/\partial v.
\tag5.12
$$
The equations \thetag{5.11} are overdetermined. Using \thetag{5.8}
and \thetag{5.12}, from \thetag{5.11} we derived the following differential
consequence:
$$
b_k\,\left(b''_q-\frac{a'}{a}\,b'_q\right)=
b_q\,\left(b''_k-\frac{a'}{a}\,b'_k\right).
\tag5.13
$$
\proclaim{Lemma 5.1} The components of two covectors $\bold b$ and
$\bold c$ satisfy the relationships $b_k\,c_q=b_q\,c_k$ if and only if
they are linear dependent.
\endproclaim
\demo{Proof} If one of the covectors is equal to zero the lemma statement is
trivial. Let $\bold b\neq 0$. Then, at least one component of the covector
$\bold b$ is not zero. Suppose $b_1\neq 0$. Then from
$b_k\,c_q=b_q\,c_k$ we derive $c_k=\lambda\,b_q$, where $\lambda=c_1/b_1$.
It means $\bold c=\lambda\,\bold b$. So, the lemma is proved.
\qed\enddemo
Let us apply Lemma 5.1 to the equation \thetag{5.13} where
$\bold b\neq 0$. So, as a consequence of \thetag{5.13} we obtain the
following equation:
$$
b''_q-\frac{a'}{a}\,b'_q=\lambda\,b_q,
\tag5.14
$$
where $\lambda$ is some scalar. Let's differentiate the equation
\thetag{5.11} with respect to $v$ and write the obtained result in the
form:
$$
b''_q-\frac{a'}{a}\,b'_q=\frac{\partial^2\ln|a|}{\partial v^2}\,
b_q+\frac{\partial^2\ln|a|}{\partial v\,\partial x^q}.
\tag5.15
$$
The comparison of \thetag{5.14} with \thetag{5.15} leads to
the equation that is also to be considered as a differential consequence
of the equations \thetag{5.7} and \thetag{5.9}:
$$
\frac{\partial}{\partial x^q}\left(\frac{a'}{a}\right)=
\mu\,b_q,
\tag5.15
$$
where the scalar $\mu$ is obtained from $\lambda$ by subtracting the
second logarithmic derivative of the function $a$ with respect to $v$.\par
      For further analysis of the obtained equations \thetag{5.15} note
that the equations \thetag{5.8} are locally solvable. Any field
$\bold b$ satisfying the equations \thetag{5.8} is defined
by some scalar field $\beta$ from the expanded algebra which
depends on the modulus of velocity only:
$$
b_r=\nabla_r\beta=\partial\beta/\partial x^r.
\tag5.16
$$
The substitution \thetag{5.16} into the equation \thetag{5.15}
transforms this equation to the form:
$$
\frac{\partial}{\partial x^q}\left(\frac{a'}{a}\right)=
\mu\,\frac{\partial\beta}{\partial x^q}.
\tag5.17
$$
\proclaim{Lemma~5.2} If spatial gradients of two functions
$\alpha(\bold x,v)$ and $\beta(\bold x,v)$ are proportional:
$\partial\alpha/\partial x^q=\mu\,\partial\beta/\partial x^q$
then in some neighborhood of points at which the gradient of
the function $\beta$ is not zero, there is the representation
$\alpha=F(\beta,v)$ where $F$ is some function of two variables.
\endproclaim
\demo{Proof} In Lemma~5.2 the quantity $v$ plays the role of a parameter.
Therefore to prove the lemma it is convenient to assume $v$ by fixed
$v=v_0=\const$. Then, while $\grad\beta\neq 0$, the local coordinates
$x^1,\ldots,x^n$ on $M$ can be taken so that $x^1=\beta(\bold x,v_0)$.
In this case from the proportionality of the gradients $\alpha$ and
$\beta$ we have:
$$
\frac{\partial\alpha}{\partial x^2}=\ldots=
\frac{\partial\alpha}{\partial x^2}=0.
$$
It means $\alpha=\alpha(x^1,v_0)$. And the functional dependence of
$\alpha$ on $x^1$ and $v_0$ in these specially chosen coordinates
defines the function $F(\beta,v)$. The choice of such coordinates can be
made smoothly depending on the parameter $v_0$. Therefore
$F(\beta,v)$ is the smooth function of two variables that defines
a relation between $\alpha$ and $\beta$ in the form of following
relationship:
$\alpha=F(\beta,v)$.
\qed\enddemo
Let's apply Lemma~5.2 to the equations \thetag{5.17}. As a result we
obtain the differential equation binding $a$ and $\beta$:
$$
\frac{\partial\ln|a|}{\partial v}=F(\beta,v).
\tag5.18
$$
Let's substitute \thetag{5.18} into the equation \thetag{5.11}. Then
$$
\frac{\partial\ln|a|}{\partial x^q}+F(\beta,v)\,
\frac{\partial\beta}{\partial x^q}=\frac{\partial^2\beta}
{\partial x^q\,\partial v}.
\tag5.19
$$
Let $\Phi(\beta,v)$ be the antiderivative of the function $F(\beta,v)$
with respect to $\beta$ for fixed $v$. In other words, the function $\Phi$
is connected with $F$ by the relationship:
$$
F(\beta,v)=\frac{\partial\Phi(\beta,v)}{\partial\beta},
\tag5.20
$$
that defines $\Phi(\beta,v)$ up to the summand depending on $v$ only:
$$
\Phi(\beta,v)\to\Phi(\beta,v)+\Psi(v).
\tag5.21
$$
After the substitution \thetag{5.20} into \thetag{5.19} taking into
account that $\beta=\beta(\bold x,v)$ lets us to rewrite the equation
\thetag{5.19} in the following form:
$$
\frac{\partial}{\partial x^q}\left(\ln|a|+\Phi(\beta,v)-
\frac{\partial\beta}{\partial v}\right)=0.
\tag5.22
$$
By virtue of \thetag{5.22} the expression in parentheses is the
quantity depending on $v$ only. Therefore because of the arbitrariness
\thetag{5.21} in the choice of function $\Phi(\beta,v)$ we have:
$$
\ln|a|=\frac{\partial\beta}{\partial v}-\Phi(\beta,v).
\tag5.23
$$
The substitution of \thetag{5.23} into the equation \thetag{5.18}
leads to the differential equation that for the given $\Phi(\beta,v)$
defines the dependence of $\beta$ on $v$:
$$
\beta''=\frac{\partial\Phi(\beta,v)}{\partial\beta}
\,(\beta'+1)+\frac{\partial\Phi(\beta,v)}{\partial v}.
\tag5.24
$$
The relation of the quantities $\beta$ and $\Phi(\beta,v)$ and
the force field of the dynamical system contains an element of arbitrariness
(see \thetag{5.21}). This arbitrariness defines the following gauge
transformations that do not change the form of the equation \thetag{5.24}:
$$
\aligned
&\beta\to\tilde\beta(\bold x,v)=\beta(\bold x,v)+\psi(v),\\
&\Phi\to\tilde\Phi(\tilde\beta,v)=\Phi(\tilde\beta-\psi(v),v)+
\psi'(v).
\endaligned
\tag5.25
$$
The transformations \thetag{5.25} do not change the quantities
$\alpha=\ln|a|$ in \thetag{5.23}. The equation \thetag{5.24} can be
rewritten in the form of a system of two equations of the first order
$$
\cases
\beta'=\alpha+\Phi(\beta,v),\\
\alpha'=\dsize\frac{\partial\Phi(\beta,v)}{\partial\beta}.
\endcases
\tag5.26
$$
Note that the system \thetag{5.26} also admits the gauge transformations
\thetag{5.25} complemented by the relationship $\alpha\to\tilde\alpha(v)=
\alpha(v)$.\par
      In general case the equation \thetag{5.24} as well as the system
\thetag{5.26} is not possible to solve explicitly. However
this equation enables to characterize exactly the extent of
arbitrariness in the definition of force field of the dynamical system
accepting the normal shift. For fixed choice of the function $\Phi(\beta,v)$
the general solution of the equation \thetag{5.24}  contains  two
integration
constants $f$ and $h$:
$$
\beta(v)=B_\Phi(v,f,h).
\tag5.27
$$
The parameters $f$ and $h$ in \thetag{5.27} can depend on the
coordinates $x^1,\ldots,x^n$. So, $f$ and $h$ are two scalar fields on
the manifold $M$: $f=f(\bold x)$ and $h=h(\bold x)$. The corresponding
force field of the dynamical system for \thetag{5.27} has the following
components:
$$
F_q=\frac{\exp(\partial B_\Phi/\partial v)}
{\exp(\Phi(B_\Phi,v))}\,N_q+
|\bold v|\,\left(\frac{\partial B_\Phi}{\partial f}\,
\nabla_if+\frac{\partial B_\Phi}{\partial h}\,\nabla_ih
\right)\,(2\,N^i\,N_q-\delta^i_q).
\tag5.28
$$
The formula \thetag{5.28} is considerably less effective than the
formulae \thetag{5.9} and \thetag{5.10} for the cases considered earlier.
It contains the function $B_\Phi$ which is not arbitrary. Some special cases
when this formula can be made more effective we'll consider in following
section.
\proclaim{Theorem~5.1} The force field $\bold F$ of the dynamical
system that admit the normal shift is locally defined by one of three
formulae \thetag{5.9}, \thetag{5.10} or \thetag{5.28}. In the last case
it contains three arbitrary parameters: two scalar fields $f(\bold x)$ and
$h(\bold x)$ and the function of two parameters $\Phi(\beta,v)$.
\endproclaim
      \subhead Example 3\endsubhead In \cite{5} and \cite{6}
the class of metrizable dynamical systems that admit the normal shift
was studied. Consider these systems in the context of the above
construction. Let us choose the function
$\Phi(\beta,v)$ of the special form:
$$
\Phi(\beta,v)=-\ln H(v\,e^{-\beta/v}),
\tag5.29
$$
where $H=H(\xi)$ is some smooth function of one variable.
The substitution of \thetag{5.29} into \thetag{5.24} leads to
the equation:
$$
\beta''=\frac{H'}{H}\,e^{-\beta/v}\,\left(\beta'
-\frac{\beta}{v}\right).
\tag5.30
$$
Finding the general solution of the equation \thetag{5.30} in
explicit form is too problematic. However one particular solution is
easily guessed:
$$
\beta(\bold x,v)=v\,f(\bold x).
\tag5.30
$$
Here $f$ is some scalar field on the manifold $M$. The substitution of
\thetag{5.29} and \thetag{5.30} into \thetag{5.23} defines the field
$a$. The field $\bold b$ is also easily calculated:
$$
\xalignat 2
&a=H(v\,e^{-f})\,e^f,
&
&b_q=v\,\nabla_qf.
\tag5.32
\endxalignat
$$
The fields \thetag{5.32} define dynamical systems with the force
field of the form:
$$
F_q=-|\bold v|^2\,\nabla_qf+2\,\nabla_kf\,v^k\,v_i+
N_q\,H(|\bold v|\,e^{-f})\,e^f.
$$
This is the force field of the metrizable dynamical systems from \cite{6}.
It contains one arbitrary scalar field $f$ and an arbitrary
function $H$ of one variable. It's clear that this case does not exhaust
the functional arbitrariness declared by theorem~5.1. This fact
confirms the existence of non-trivial non-metrizable dynamical systems
accepting the normal shift in the dimension $n\geq 3$.\par
\head
6. Some new examples.
\endhead
\subhead Example 4 \endsubhead Consider the case, when the equation
\thetag{5.24} becomes linear. So, let us choose the function
$\Phi(\beta,v)$ being linear with respect to $\beta$ and set
$$
\Phi(\beta,v)=-\frac{\phi''(v)}{\phi'(v)}\,\beta+\ln\phi'(v).
\tag6.1
$$
The expression \thetag{6.1} is not a general form of the function linear
in $\beta$, however the general case can be reduced to \thetag{6.1}
by the gauge transformation \thetag{5.25}. The system of
equations \thetag{5.6} equivalent to the equation \thetag{5.24}
for the function $\Phi(\beta,v)$ of the form \thetag{6.1} is written as:
$$
\cases
\beta'=\alpha-\frac{\phi''(v)}{\phi'(v)}\,\beta+\ln\phi'(v),\\
\alpha'=-\frac{\phi''(v)}{\phi'(v)}
\endcases
\tag6.2
$$
The system \thetag{6.2} is integrated easily. As constants of
integration two scalar fields $f(\bold x)$ and $h(\bold x)$
arise:
$$
\xalignat 2
&\alpha=-\ln\phi'(v)+\ln h,
&
&\beta=\frac{\ln h\,\phi(v)+f}{\phi'(v)}.
\tag6.3
\endxalignat
$$
Hence for the force field $\bold F$ of the dynamical system
by virtue of \thetag{4.20} and the scalar substitution \thetag{3.2}
we obtain the expression:
$$
F_q=\frac{h}{\phi'(|\bold v|)}\,N_q+|\bold v|\,\left(
\frac{\nabla_ih}{h}\,\frac{\phi(|\bold v|)}{\phi'(|\bold v|)}
+\frac{\nabla_if}{\phi'(|\bold v|)}\right)\,
(2\,N^i\,N_q-\delta^i_q).
\tag6.4
$$\par
      Consider the equation \thetag{5.24} once again. The general solution
of this equation depends on two integration constants: $\beta=B_\Phi(v,f,h)$.
Let the parameter $h$ be functionally expressed via the parameter $f$, i.e.
$h=h(f)$. This diminishes the number of arbitrary parameters and leads to the
function
$$
\beta=\beta(v,f)=B_\Phi(v,f,h(f)),
\tag6.5
$$
that defines some one-parameter subfamily in two-parameter
family of solutions of the equation \thetag{5.24}. Let's consider
\thetag{6.5} as a function of two variables and differentiate it with
respect to $v$. As a result we obtain one more function of two variables:
$\beta'=\beta'(v,f)$. Without loss of generality, one can suppose that
the dependence on $f$ in \thetag{6.5} is locally reversible: $f=f(\beta,v)$.
Let's substitute this into the function $\beta'=\beta'(v,f)$. As a result
we obtain the following differential equation of the first order:
$$
\beta'=U(\beta,v),
\tag6.6
$$
where $U(\beta,v)=\beta'(v,f(\beta,v))$. The equation \thetag{6.6} is
compatible  with  \thetag{5.24}.  One-parameter  family  of   its
solutions
is exactly the subfamily \thetag{6.5} of solutions of the equation
\thetag{5.24}. Let us differentiate \thetag{6.6} with respect to $v$
and substitute the result into \thetag{5.24}. It leads to the relationship:
$$
\frac{\partial U(\beta,v)}{\partial\beta}\,U(\beta,v)+
\frac{\partial U(\beta,v)}{\partial v}=
\frac{\partial\Phi(\beta,v)}{\partial\beta}
\,(U(\beta,v)+1)+\frac{\partial\Phi(\beta,v)}{\partial v},
\tag6.7
$$
that binds the functions $U(\beta,v)$ and $\Phi(\beta,v)$.
The relationship \thetag{6.7} is the compatibility condition of two
ordinary differential equations \thetag{6.6} and \thetag{5.24}.\par
      Note that \thetag{6.7} is one equation for two functions. To solve
\thetag{6.7} let's introduce a new function
$W(\beta,v)=U(\beta,v)-\Phi(\beta,v)$. Then the equation \thetag{6.7}
can be written in the form:
$$
\frac{\partial U(\beta,v)}{\partial\beta}-
\frac{\partial W(\beta,v)}{\partial\beta}\,U(\beta,v)=
\frac{\partial W(\beta,v)}{\partial v}+
\frac{\partial W(\beta,v)}{\partial\beta}.
\tag6.8
$$
Let's make the following substitution into the differential equation
\thetag{6.8}:
$$
\xalignat 2
 &W(\beta,v)=-\ln w(\beta,v),
&&U(\beta,v)=\frac{u(\beta,v)}{w(\beta,v)}.
\tag6.9
\endxalignat
$$
After this substitution the equation \thetag{6.8} becomes quite
simple:
$$
\frac{\partial u}{\partial\beta}+
\frac{\partial w}{\partial\beta}+
\frac{\partial w}{\partial v}=0.
\tag6.10
$$
The general solution of the equation \thetag{6.10} is defined locally
by one arbitrary function of two variables $\phi(\beta,v)$:
$$
\xalignat 2
&u=-\frac{\partial\phi}{\partial\beta}-
\frac{\partial\phi}{\partial v},
&&w=\frac{\partial\phi}{\partial\beta}.
\endxalignat
$$
Hence for the function $U(\beta,v)$ and for the function $W(\beta,v)$
introduced above we obtain:
$$
\xalignat 2
&U=-\frac{\partial\phi/\partial v}{\partial\phi/\partial\beta}-1,
&&W=-\ln\left(\frac{\partial\phi}{\partial\beta}\right).
\tag6.11
\endxalignat
$$
Let us write the first relationship from \thetag{6.11} in the form of
an equation for the function $\phi(\beta,v)$:
$$
\frac{\partial\phi(\beta,v)}{\partial\beta}\,(U(\beta,v)+1)+
\frac{\partial\phi(\beta,v)}{\partial v}=0.
\tag6.12
$$
Such equation is solved by the methods of characteristics.
The characteristics of the equation \thetag{6.12} are defined by
the following ordinary differential equation
$$
\beta'=U(\beta,v)+1,
\tag6.13
$$
and the function $\phi(\beta,v)$ by virtue of \thetag{6.12}
is the first integral (conservation law) of the equation \thetag{6.13}.
Hence a value of the function $\phi$ can be used to parameterize the
family of solutions of the equation \thetag{6.13}. Let's write it as:
$$
\beta=B(v,\phi).
\tag6.14
$$
The function \thetag{6.14} differs from \thetag{6.5}. It inverts
the functional dependence $\phi$ on $\beta$, i.e.
$B(v,\phi(\beta,v))\equiv\beta$. Differentiating this relationship
we obtain:
$$
\xalignat 2
&\frac{\partial B}{\partial\phi}\,
 \frac{\partial\phi}{\partial\beta}=1,
&
&\frac{\partial B}{\partial v}+
\frac{\partial B}{\partial\phi}\,
 \frac{\partial\phi}{\partial v}=0.
\tag6.15
\endxalignat
$$
Let us express the derivatives $\partial\phi/\partial\beta$
and $\partial\phi/\partial v$ from \thetag{6.15} via
$\partial B/\partial\phi$ and $\partial B/\partial v$ and
substitute the result into \thetag{6.11}. It gives:
$$
\xalignat 2
&U=\frac{\partial B}{\partial v}-1,
&
&W=\ln\left(\frac{\partial B}{\partial\phi}\right).
\tag6.16
\endxalignat
$$
Form \thetag{6.16} it is easy to define the function $\Phi$ but
in the variables $v$ and $\phi$:
$$
\Phi=U-W=B_v-\ln(B_\phi)-1.
\tag6.17
$$
In the same variables $v$ and $\phi$ it is naturally to write the equation
\thetag{6.6} too. So, we obtain the equation defining the dependence
of $\phi$ on $v$:
$$
\phi'=-\frac{1}{B_\phi(v,\phi)}.
\tag6.18
$$
It is derived from $\phi'=\partial\phi/\partial\beta\cdot U+\partial
\phi/\partial v$ and from the relationships \thetag{6.15} defining
partial derivatives $\phi$ with respect to $v$ and $\beta$.\par
      Summarizing the computations just done, note that the initial
equation \thetag{5.24} contains an arbitrary function $\Phi(\beta,v)$,
we've expressed it via $\phi(\beta,v)$ and then via the function
$B(v,\phi)$. Hence it is convenient to consider $B(v,\phi)$ as
a primary function and to express everything in terms of this function.
      \subhead Example 5 \endsubhead Let us render concrete the choice
of the function $B(v,\phi)$ to solve the equation \thetag{6.18} explicitly.
Let
$$
B(v,\phi)=-\frac{1}{3\,v}\,\ln\left(\frac{\phi}{\phi+3}\right).
$$
Then the equation \thetag{6.18} for the function $\phi(v)$ gets
the form of the Bernoulli equation:
$$
\phi'=\phi^2\,v+3\,v\,\phi.
$$
The general solution of this equation contains one integration constant
$f$ that depends on the space variables $x^1,\ldots,x^n$ and defines
a scalar field on the manifold $M$:
$$
\phi=\phi(v,f)=\frac{3}{\exp(-3\/v^2/2-2\,f)-1}.
$$
Now it remains to compute the fields $a$ and $\bold b$ in \thetag{4.20}
to define the field $A$ of the scalar substitution \thetag{3.2}
$$
\aligned
&a=e^W=B_\phi=-\frac{4}{9\,v}\,
\sinh^2\left(-\frac{3\,v^2}{4}-f\right),\\
\vspace{1ex}
&b_i=\frac{\partial\beta}{\partial x^i}=
\frac{\partial B}{\partial\phi}\,\frac{\partial\phi}{\partial f}\,
\nabla_if=-\frac{\nabla_if}{3\,v}.
\endaligned
$$
Now the expression for the components of force field of the dynamical
system accepting the normal shift is easily derived:
$$
F_q=-\frac{4}{9\,|\bold v|}\,\sinh^2\left(-\frac{3\,|\bold v|^2}
{4}-f\right)\,N_q-\frac{\nabla_if\,(2\,N^i\,N_q-\delta^i_q)}{3}.
\tag6.19
$$
The force field \thetag{6.19} contains one scalar field $f$ only
and this fact distinguishes this force field from the general case
\thetag{5.28} and from the case \thetag{6.4}. It happens because we've
replaced the equation of the second order \thetag{5.24} by the equation of
the first order \thetag{6.6}.\par
\head
8. Final remarks.
\endhead
      Three cases defined by the formulae \thetag{5.9}, \thetag{5.10} and
\thetag{5.28} characterize completely the local construction of the
dynamical systems accepting the normal shift on Riemannian manifolds
of the dimension $n\geq 3$. Two dimensional case is distinguished by a
larger extent of diversity. In multidimensional case the system of normality
equations is strongly overdetermined (however it is compatible). Hence the
set of its solutions is more restricted.\par
      Note that the existence and number of local solutions
of normality equations are not connected with the features of metric:
the constantness of curvature, the number of isometries etc. Hence the
existence and number of smooth global solutions of these equations
depend exclusively on the topology of the manifold $M$.\par

\enddocument
\end